\documentclass[
  ,final            
,numberedheadings 
  ]
  {aipproc}

\layoutstyle{6x9}
\usepackage{amssymb,amsmath}
\usepackage{wrapfig}
\usepackage{macros,macros_static}

\begin{document}

\title{Non-perturbative Heavy Quark Effective Theory:
a test and its matching to QCD 
\footnote{Invited talk at ``6th Conference on Quark Confinement and the 
	Hadron Spectrum'', Sept. 21--25 2004, Villasimius, Italy}
\vspace{-3cm}
\begin{flushright}
\normalsize
DESY 04-246 \hspace{0.5cm}
SFB/CPP-04-71 \hfill \today
\end{flushright}
\vspace {2cm}
}


\author{Rainer Sommer}{
  address={DESY, Platanenallee 6, 15738 Zeuthen, Germany}
}

\begin{abstract}
  We give an introduction to the special problems encountered
  in a treatment of HQET beyond perturbation theory in the 
  gauge coupling constant.
  In particular, we report on a recent test of HQET 
  as an effective theory for QCD and
  discuss how HQET can be implemented on the lattice 
  including the non-perturbative matching of the effective theory to QCD. 
\end{abstract}

\maketitle


\section{Introduction }
\label{s:intro}

Heavy quark effective theory is routinely used
in phenomenology. In these applications, the matching to QCD is 
achieved perturbatively and matrix elements of the 
operators in the effective theory are determined from
experiment and models. However, HQET took its origin as an effective
theory in the lattice regularization, where it was designed
as a solution to the problem of treating quarks which
are heavy compared to the inverse lattice spacing and
thus do not propagate properly in the standard relativistic
framework~\cite{stat:eichten}. 

Unfortunately, after considerable initial activity (see e.g. 
\cite{reviews:eichten90,reviews:beauty,reviews:hartmut97} and
references therein)
the non-perturbative treatment of the effective theory on the 
lattice had been somewhat dormant for a while. The reason is
that it was realized \cite{stat:MaMaSa} that a non-perturbative
matching to QCD is needed; otherwise the continuum
limit does not exist. A practicable solution of this
problem was only found recently \cite{zastat:pap3,lat01:rainer,hqet:pap1}. 

Here we point out that a non-perturbative matching is
necessary on {\em and off} the lattice, 
the problem being most severe on the lattice
and we review a non-perturbative test of HQET.
We then explain a recent strategy to perform fully
non-perturbative computations in HQET  
and discuss the status and perspectives of this 
approach.

\section{HQET as an asymptotic expansion of QCD} \label{s:asy}

Consider QCD at energies low enough such that the 
top-quark may be neglected altogether. In the QCD
Lagrangian
\bea
\lag{\rm QCD}=-{1 \over 2g_0^2}\tr \{F_{\mu\nu}F_{\mu\nu}\}
                    +\sum_f \psibar_f [D_\mu\gamma_\mu +m_f ]\psi_f
\eea
the sum over flavors then extends over $f={\rm u,d,s,c,b}$. 
An effective theory, HQET
is expected to provide the asymptotic expansion of 
a certain (large) set of energies (e.g. mass splittings) and matrix elements of QCD
in terms of the inverse of the mass of the b-quark.\footnote{
	Powers of $1/\mbeauty$ are understood to be accompanied
	by slowly (logarithmically) varying functions of $\mbeauty$.}
We restrict our discussion to the energies and matrix elements
of states which contain a single b-quark at rest
and refer to reviews such as \cite{HQET:neubert,reviews:HQETMannel} for 
the general case of finite velocity.
The Lagrangian of the theory, which may be obtained by
a formal $1/\mbeauty$ expansion (see e.g. \cite{nrqcd:first}), is then
given by the replacement
\bea \label{e:statlag}
    \psibar_\mrm{b} [D_\mu\gamma_\mu + \mbeauty ]\psi_\mrm{b}
     &\to& \lag{\rm stat}+\lnu{1} + \ldots 
	\,,\quad
    \lag{\rm stat}=\heavyb [D_0 + \dmstat ]\heavy \,.
\eea
Here $\lnu{1}$ is of order $1/\mbeauty$
and the mass term of the b-quark has been removed from the Lagrangian
such that observable quantities in the b-sector have a finite limit as 
$ \mbeauty\to\infty$ (with a suitable counter term $\dmstat$).
The effective heavy quark field $\heavy$ has only two degrees of freedom 
as appropriate for a non-relativistic spin 1/2 particle. Still it is
notationally convenient to keep $\heavy$ as a 4-component spinor but impose
the constraint
\bea                                
    \qquad P_+\heavy=\heavy\,,\;P_+=(1+\gamma_0)/2 \,;
\eea
i.e. the lower components vanish in the Dirac representation. 
In order to discuss matrix elements, such as
the B-meson decay constant, also the composite fields involving
b-quarks are translated to the effective theory, for example:
\bea  \label{e:astat}                              
   A_0 = \za \psibar_\mrm{l} \gamma_0\gamma_5 \psi_{\rm b} \quad
    &\to&\quad \Astat = \zastat \psibar_\mrm{l} \gamma_0 \gamma_5 \heavy \,.
\eea
Here $\za,\zastat$ are 
the renormalization constants of the composite fields. 

The effective theory is valid for the low-lying 
energy levels as well as their matrix elements, the simplest one
being
\bea
      \Phiqcd \equiv \fb\sqrt{\mb}=\za \langle0| A_0 |B \rangle\,.
\eea
It is scale independent, due to the chiral symmetry of QCD
in the massless limit (including $\mbeauty=0$). In the effective
theory this symmetry is absent and $\zastat$ depends on the energy
scale, $\mu$, used in the renormalization condition which defines the
finite current. 
Instead of 
$\Phistat(\mu)\equiv\zastat(\mu) \langle0| A_0 |B \rangle_\mrm{stat}$ 
it is therefore better to consider the 
renormalization group invariant matrix element
\bes \label{e:phirgi}
  \PhiRGI
                   = \lim_{\mu\to\infty} \left[\,2b_0 \gbar^2(\mu)\,\right]^{-\gamma_0/2b_0}
                   \Phistat(\mu)\,.
\ees
It is both $\mu$ and renormalization scheme independent, as is easily 
seen using $\Phistat_\mrm{scheme}(\mu) =\Phistat_\mrm{scheme'}(\mu)(1+\rmO(\gbar^2(\mu))$.
In \eq{e:phirgi}, 
the coefficients $b_0,\gamma_0$  defined by 
\bes
   \beta(\gbar) \equiv \mu {\rmd \over \rmd \mu} \gbar  = - b_0 \gbar^3 + \rmO(\gbar^5)\,,\quad 
   \gamma(\gbar) \equiv  {\mu \over  \zastat} {\rmd \over \rmd \mu} \zastat 
   = - \gamma_0 \gbar^2 + \rmO(\gbar^4)
\ees
enter. We can now write down
the HQET-expansion of the QCD matrix element 
\bes \label{e:matchPhi}
  \Phiqcd &=& \Cps(M_\beauty/\Lambda_\msbar)\times
                   \PhiRGI\,+\, \rmO(1/M_\beauty) \, , \\
  M_\beauty &=& \lim_{\mu\to\infty} \left[\,2b_0 \gbar(\mu)\,\right]^{-d_0/2b_0}
                   \mbar(\mu)\,, \quad    \tau(\gbar) \equiv  {\mu \over  \mbar} {\rmd \over \rmd \mu} \mbar 
   = - d_0 \gbar^2 + ... \\
   \label{e:lambda}
  \Lambda_\msbar &=&  
	\lim_{\mu\to\infty}  \mu\ \left(b_0\gbarmsbar^2(\mu)\right)^{-b_1/(2b_0^2)}\
  \rme^{-1/(2b_0\gbarmsbar^2(\mu))} \,.
\eea
Let us dicuss the somewhat unfamiliar form of \eq{e:matchPhi} 
and the conversion function  $\Cps(M_\beauty/\Lambda_\msbar)$.
In a more conventional form we have
\bea \label{e:match1}
   \Phiqcd &=&C_\mrm{match}(\mbeauty/\mu)\times\PhiMSbar(\mu)
                                + \rmO(1/\mbeauty)
\eea
with a matrix element $\PhiMSbar(\mu)$ renormalized in the effective
theory in the $\MSbar$ scheme and the matching coefficient
$C_\mrm{match}(\mbeauty/\mu)$ depending on the b-quark mass $\mbeauty$ in
the $\MSbar$ scheme at scale $\mbeauty$, i.e. $\mbar_\MSbar(\mbeauty)=\mbeauty$. 
The factor $C_\mrm{match}$ is determined (usually in perturbation theory)
such that \eq{e:match1} holds for some particular matrix element
of the current and will then be valid {\em for all matrix elements}.
Contact to \eq{e:matchPhi} is easily made by using 
\bea
   {\PhiRGI \over \PhiMSbar(\mu)} = \left[\,2b_0 \gbar^2(\mu)\,\right]^{-\gamma_0/2b_0}
                   \exp\left\{-\int_0^{\gbar(\mu)} \rmd g
                     \left[\,{ \gamma_{\MSbar}(g) \over\beta_{\MSbar}(g)}
		           -{\gamma_0 \over b_0 g}\,\right]
                     \right\} \,,
\eea
setting the arbitrary renormalization point $\mu$ 
to $\mbeauty$ and identifying 
\bea \label{e:cps}
      \Cps\left({M_\beauty \over\Lambda_\MSbar}\right) &=&
                  C_\mrm{match}(1){\PhiMSbar(\mbeauty) \over \PhiRGI} \\
    &=& \left[\,2b_0 \gbar^2(\mbeauty)\,\right]^{\gamma_0/2b_0}
                   \exp\left\{\int_0^{\gbar(\mbeauty)} \rmd g
                     \left[\,{ \gamma_\mrm{match}(g) \over\beta_{\MSbar}(g)}
		           -{\gamma_0 \over b_0 g}\,\right]  \right\} \,, \nonumber
\eea
where $\gbar$ is taken in the $\msbar$ scheme.
The last equation may be taken as a definition of the 
anomalous dimension $\gamma_\mrm{match}$ in the ``matching scheme''. 
It has contributions from $\gamma_{\MSbar}$ as well as from  $C_\mrm{match} = 1 + c_1 \gbar^2 +\ldots$, namely 
\bes \label{e:gammamatch}
   \gamma_\mrm{match}(\gbar) = -  \gamma_0\gbar^2 - [\gamma_1^{\msbar} + 2b_0 c_1]\gbar^4
   + \ldots \,.
\ees
Note that replacing the $\msbar$ coupling by a non-perturbative one, 
$\gamma_\mrm{match}$ may also be defined beyond perturbation theory 
through eqs.~(\ref{e:cps},\ref{e:matchPhi}).\footnote{Clearly the r.h.s.
of \eq{e:cps} is a function of $\gbar^2(\mbeauty)$, i.e. a function
of $\mbeauty/\Lambda_\msbar$. We prefer to write it as a function
of the ratio of renormalization group invariants, $M_\beauty /\Lambda_\MSbar$.}
Another advantage of \eq{e:cps} 
is that $\Cps$ is independent of the arbitrary choice of renormalization
scheme for the effective operators in the effective theory. Apart from the 
choice of the QCD coupling, the ``convergence'' of the series
\eq{e:gammamatch} is dictated by the physics, nothing else.
Note further that (at leading order in $1/M$) the conversion function
$\Cps$ contains the full (logarithmic) mass-dependence. 
The non-perturbative effective theory matrix elements are mass independent
numbers.
Conversion functions such as $\Cps$ are universal for all 
(low energy) matrix 
elements of their associated operator. Thus
\bea
        \za^2 \langle A_0^\dagger(x) A_0(0)\rangle_\mrm{QCD}
                    &\simas{x^2 \gg 1/M_\beauty^2}&
                    [\Cps(\frac{M_\beauty}{\Lambda_\msbar})]^2
                    \langle \Astat(x)^\dagger \Astat(0)\rangle_\mrm{RGI} +\rmO(\frac{1}{M_\beauty})\,
\eea
is a straight forward generalization of \eq{e:matchPhi}.

\begin{wrapfigure}{r}{7.0cm}
  \hspace{0.5cm}\includegraphics[width=6.5cm]{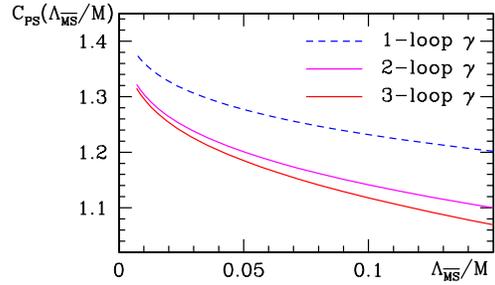}
  \caption{\footnotesize $\Cps$ estimated in perturbation
	theory.  
	}\label{f:cps}
\end{wrapfigure}

Analogous expressions for the conversion functions
are valid for the time component of the 
axial current replaced by other composite fields, for example 
the space components of the vector current. 
Based on the work of \cite{BroadhGrozin1,Shifman:1987sm,Politzer:1988wp}
and recent efforts
their perturbative expansion is known including the 3-loop anomalous dimension 
$\gamma_\mrm{match}$ obtained from the 
3-loop anomalous dimension  $\gamma_{\MSbar}$ \cite{ChetGrozin} 
and the 2-loop matching function $C_\mrm{match}$ 
\cite{Ji:1991pr,BroadhGrozin2,Gimenez:1992bf}.
\Fig{f:cps}, taken from \cite{hqet:pap3}, 
illustrates that the remaining $\rmO(\gbar^6(\mbeauty))$ errors in $\Cps$ 
seem to be relatively small. 

Although it is generally accepted that HQET is an effective theory of
QCD in the sense that was just described, tests of this equivalence
are rare and mostly based on phenomenological analysis of experimental
results. A pure theory test can be performed if
QCD including a heavy enough quark can be simulated on the lattice
at lattice spacings which are small enough to be able to take the continuum limit. 
This has recently been achieved~\cite{hqet:pap3}
and will be summarized below.

\subsection{Tests of HQET in a finite volume} \label{s:tests}

\begin{wrapfigure}{r}{6.0cm}
  \hspace{0.5cm}\includegraphics[width=5.5cm]{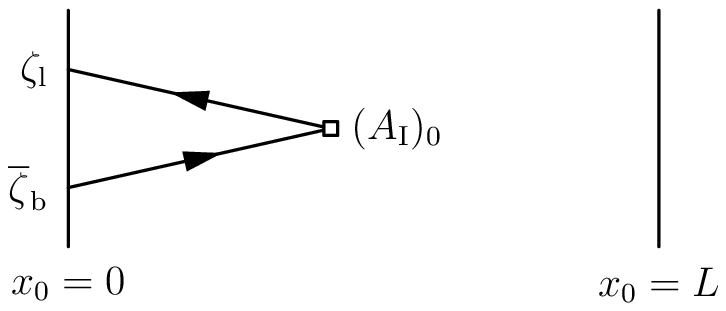}
\caption{ \footnotesize
The correlation function $\fa$. 
}\label{f:correlators}
\end{wrapfigure}
We start with the QCD side of such a test. 
Lattice spacings such that $a \mbeauty \ll 1$ can be reached if 
one puts the theory in a finite volume, $L^3 \times T$ with $L,T$ not
too large. We shall use $T=L$. 
For various practical reasons, so-called
\SF boundary conditions are chosen, i.e. Dirichlet in time
(at $x_0=0,T)$
and periodic in space \cite{SF:LNWW,SF:stefan1}. Equivalent
boundary conditions are easily imposed in the effective theory 
\cite{zastat:pap1}. We then form correlation functions
of boundary quark fields $\zeta$ (located at $x_0=0$) and composite fields
such as the time component of the axial current in the bulk ($0<x_0<T$),
as illustrated in 
\fig{f:correlators} and given for example by
\bes  \label{e:fa}
  \fa(x_0) &=& -{a^6 \over 2}\sum_{\vecy,\vecz}\,
  \left\langle
  (\aimpr)_0(x)\,\zetabar_{\rm b}(\vecy)\gamma_5\zeta_{\rm l}(\vecz)
  \right\rangle  \,. 
\ees
(The current
$\aimpr$ represents the $\rmO(a)$-improved version of the
axial current for which lattice artifacts linear in the lattice
spacing are absent.)
Another correlation function, $\fone$, describes the propagation of
a quark-antiquark pair from the $x_0=0$ boundary to the 
 $x_0=T$ boundary. 
For details we refer to \cite{hqet:pap3}. 

We then take a ratio for which the renormalization
factors of the boundary fields cancel,
\bes
\Yr(L,M) &\equiv& \za \left.{\fa(L/2) \over \sqrt{f_1}}\right|_{T=L} =
            {\langle \Omega(L)| A_0 |B(L)\rangle
              \over
              || \,| \Omega(L)\rangle \,||\; ||\, | B(L)\rangle \, ||}
            ,\\
            && |B(L)\rangle = \rme^{ -L H /2 } |\varphi_{\rm B}(L)\rangle \,,\;
            |\Omega(L)\rangle =\rme^{ -L H/2 } |\varphi_{0}(L)\rangle \,.
\ees
As shown in the above equations, $\Yr$ can be represented as
a matrix element of the axial current between a 
normalized state $ |B(L)\rangle$ 
with the quantum numbers of a B-meson and  $|\Omega(L)\rangle$ which
has vacuum quantum numbers. The time evolution $ \rme^{-LH /2 }$ ensures
that both of these states are dominated by energy eigenstates with energies
around $2/L$ and less. In other words, HQET is applicable if $1/L \ll M$ (and of course 
$\Lambda \ll M$).

One then expects (for fixed $L \Lambda$) 
\bes \label{e:yrequiv}
	\Yr(L,M) / \Cps(M/\Lambda) = \XRGI + \rmO(1/z)\,, \quad z=ML\,,
\ees
where the $1/M$ corrections are written in the dimensionless
variable $1/z$ and $\XRGI$ is defined as $\Yr$ but 
at lowest order in the effective theory and renormalized as in \eq{e:phirgi}. 
\begin{figure}[tb]
\centering
  \includegraphics[height=.22\textheight]{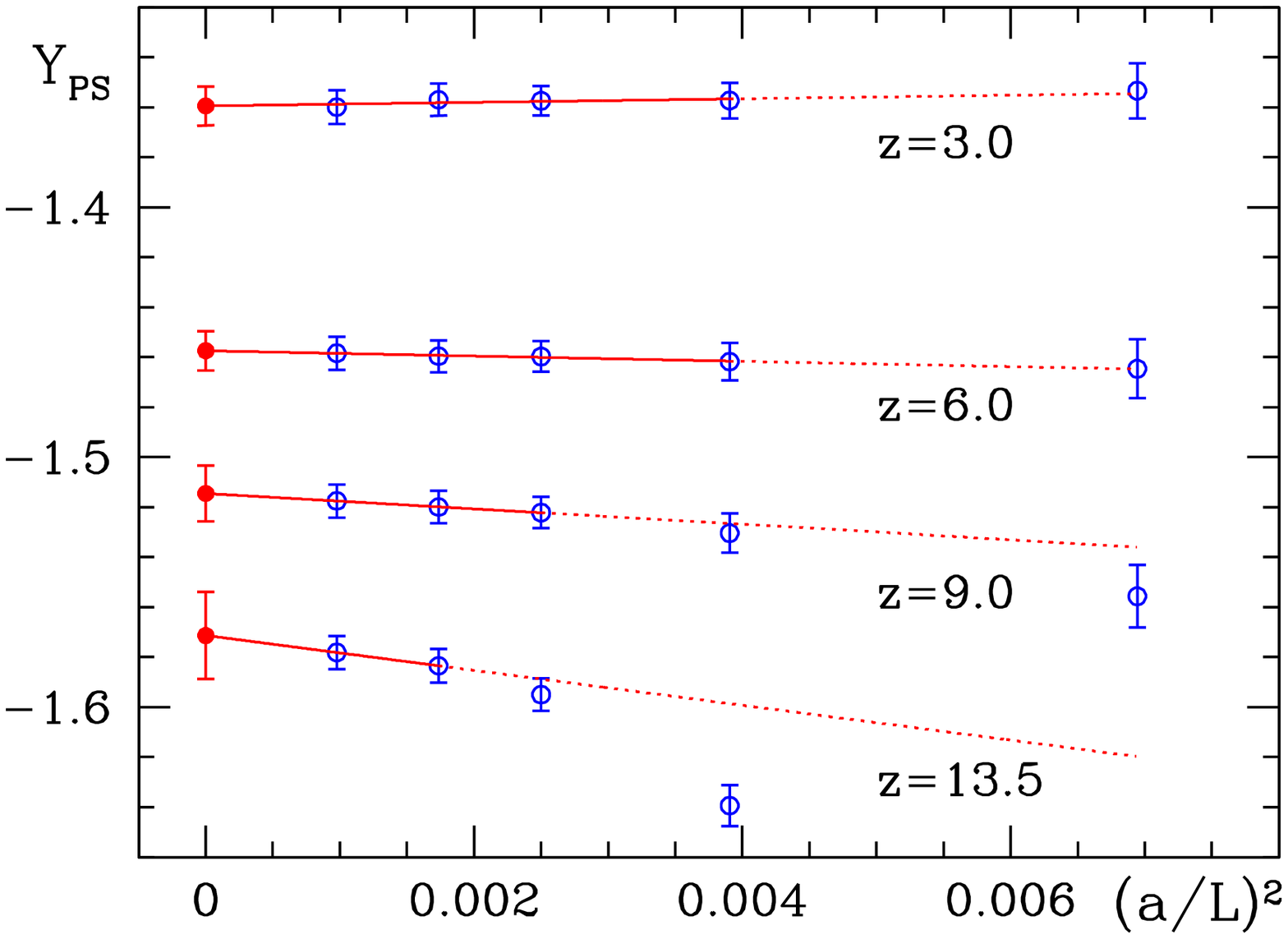}\hspace{0.2cm}
  \includegraphics[height=.22\textheight]{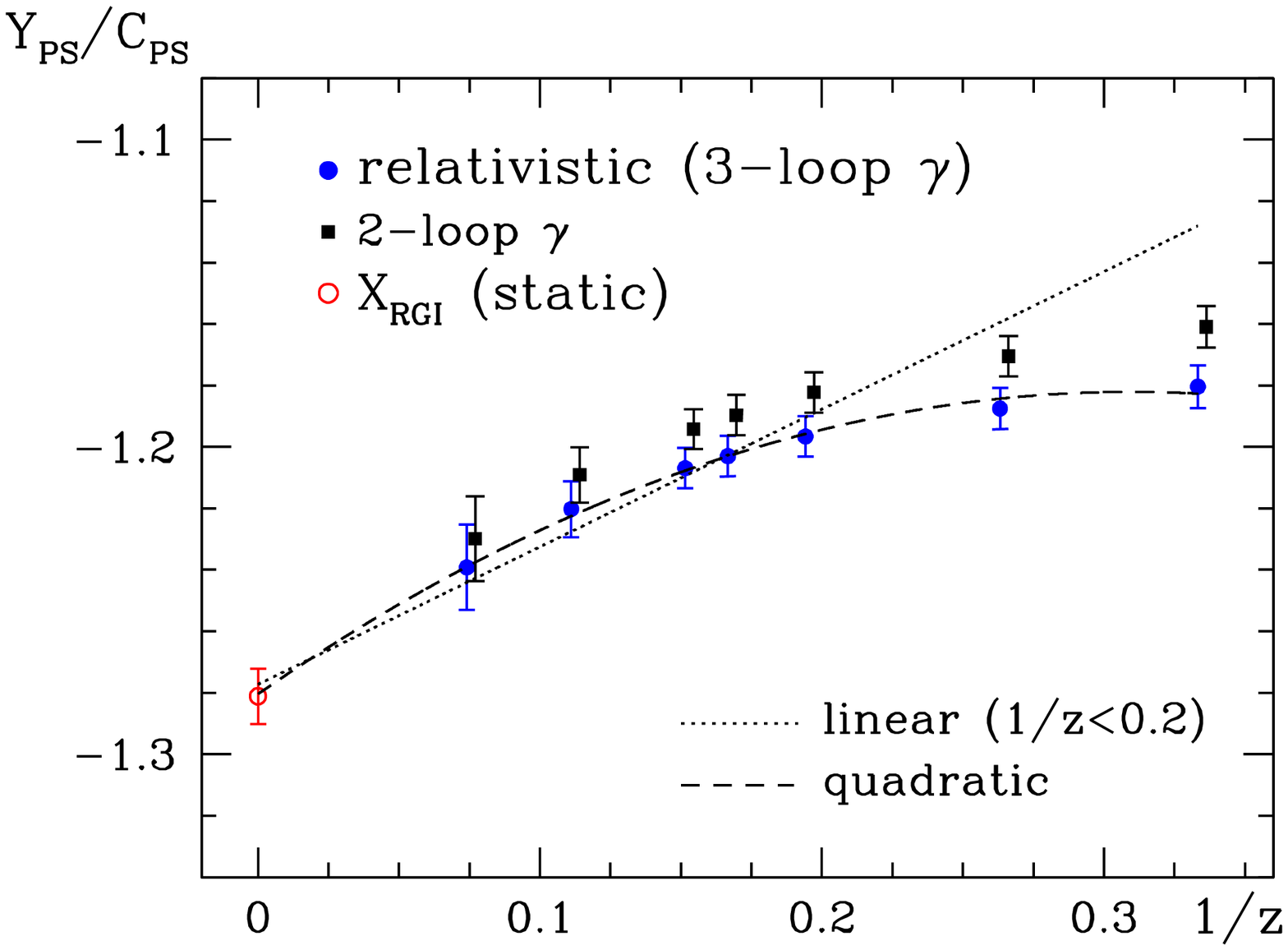}
\caption{ \footnotesize
Testing \eq{e:yrequiv} through numerical simulations in
the quenched approximation and for $L\approx0.2\,\fm$ \protect\cite{hqet:pap3}.
The physical mass of the b-quark corresponds to $z\approx5$. 
}\label{f:yrmatch}
\end{figure}
Of course such relations are expected after the continuum limit 
of both sides have been taken separately. For the case of 
$\Yr(L,M)$, this is done by the following steps:
\begin{itemize}
\item Fix a value $u_0$ for the renormalized coupling 
	$\gbar^2(L)$ (in the \SF scheme) at
	vanishing quark mass. In \cite{hqet:pap3}  
	$u_0$ is chosen such that 
	$L\approx0.2\fm$. 
\item For a given resolution $L/a$, determine the bare coupling
	from the condition $\gbar^2(L)=u_0$. This can easily be done since
	the relation between bare and renormalized coupling is known~\cite{mbar:pap1}. 
\item Fix the bare quark mass $\mqtilde$ of the heavy quark such that
	$LM=z$  using the known renormalization
	factor $\zm$ in $M =\zm \mqtilde $~\cite{mbar:pap1}.
\item Evaluate $\Yr$ and repeat for better resolution $a/L$.
\item Extrapolate to the continuum as shown in \fig{f:yrmatch}, left.
\end{itemize}
%
\begin{wrapfigure}{r}{6.5cm}
 \vspace{-0.6cm}
  \hspace{0.5cm}\includegraphics[width=6.0cm]{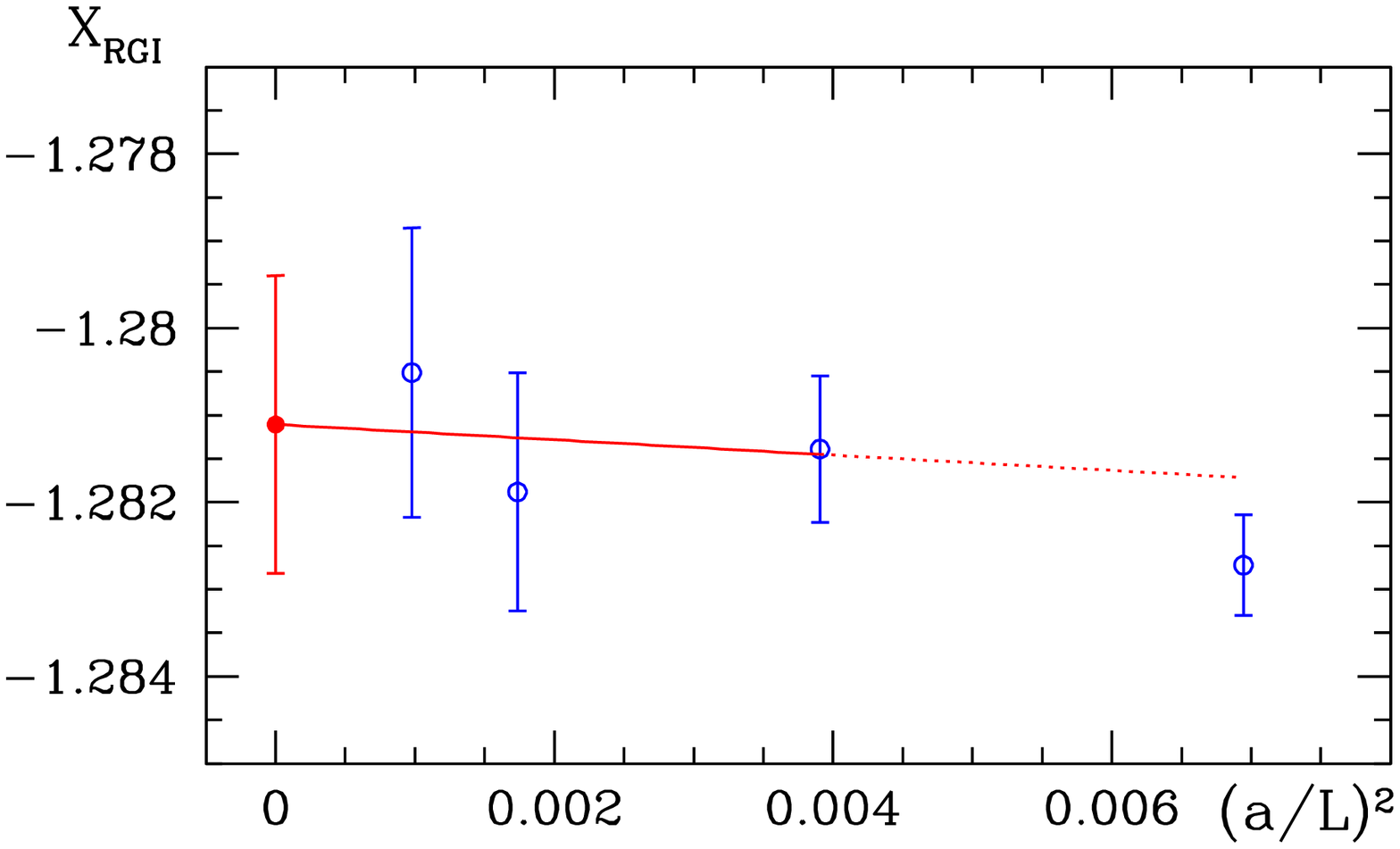}\hspace{0.0cm}
 \vspace{-0.7cm}
\caption{
\footnotesize
Continuum extrapolation of $\XRGI$ \protect\cite{hqet:pap3}.
}
\label{f:XRGI}
\end{wrapfigure}As can be seen in the figure, the continuum extrapolation becomes
more difficult as the mass of the heavy 
quark is increased and $\rmO((aM)^2)$ discretization errors
become more and more important. In contrast the continuum 
extrapolation in the static effective theory (\fig{f:XRGI}) is much easier
(once the renormalization factor relating bare current and RGI current
is known~\cite{zastat:pap3}). After the continuum limit
has been taken, the finite mass QCD observable $\Yr(L,M)$ turns
smoothly into the prediction from the effective theory
as illustrated in the r.h.s. figure. Indeed, several such successful
tests were performed in \cite{hqet:pap3},
one of them free of the perturbative uncertainty in
the conversion function (due to reparametrization 
invariance~\cite{HQET:reparaI,HQET:reparaII,HQET:reparaIII})
and two others with the static ($M\to\infty$) limit known from
the spin symmetry of HQET. For lack of space we do 
not show more examples. We only note that the coefficient
of the $1/z^n$ terms in naive fits to the finite mass results are roughly
of order unity.

Of course, finite mass lattice QCD results have been compared to 
the static limit over the years, see for example 
\cite{fb_wupp,reviews:beauty,reviews:hartmut97,El-Khadra:1998hq,
Aoki:1998ji,Bernard:1998xi,heavylight:Bec98,AliKhan:2000eg,Bowler:2000xw,
Lellouch:2000tw,lat01:ryan,mb:roma2,fb:roma2c}
and references therein. The new
quality of the tests just discussed is that the composite 
fields were renormalized non-perturbatively throughout
and that, by considering a small volume, the continuum limit could be taken
at large quark masses. 

\subsection{Beyond the leading order: the need for non-perturbative
	conversion functions} \label{s:need}

Both from looking at \fig{f:yrmatch} and from just a naive estimate 
of $\Lambda/\Mbeauty$, one expects that the effective theory has to
be implemented beyond the leading order in $1/M$ 
to reach an acceptable precision in this expansion. However, if
one wants to do this consistently, i.e. one wants to obtain
the true coefficients in the expansion, the leading
order conversion functions such as  $\Cps$ have to be known
non-perturbatively. This general problem in the 
determination of power corrections in QCD is seen by considering 
the error made in \eq{e:cps} (or \eq{e:match1}) when the 
anomalous dimension has been computed at $l$ loops and 
$C_\mrm{match}$ at $l-1$ loop order. The conversion
function $\Cps$ is then known up to an error 
\bes \label{e:deltacps}
     \Delta(\Cps) \propto  [\gbar^2(\mbeauty)]^{l} \sim
        \left\{{1 \over 2b_0\ln(\Mbeauty/\Lambda)}\right\}^{l}
        \ggas{\Mbeauty\to\infty}\; {\Lambda \over \Mbeauty} \,.
\ees
As $\mbeauty$ is made large, this error becomes dominant. Taking a 
perturbative conversion function and adding power corrections
to the leading order effective theory is thus to be regarded
as a phenomenological approach, where one assumes
that the coefficient of the $[\gbar^2(\mbeauty)]^{l}$ term 
is small, such that the $(\Lambda/\Mbeauty)^n$ corrections dominate over a certain
mass interval. Indeed, returning to our example, \fig{f:yrmatch} indicates 
that the power corrections are larger than the 
perturbative ones at $1/z=0.1 \ldots 0.2$.
Nevertheless, it remains a  fact that a theoretically 
consistent evaluation of power corrections requires 
a fully non-perturbative formulation of the theory including
a non-perturbative matching to QCD.

\section{Non-perturbative formulation of HQET} \label{s:form} 

The discussion in this section summarizes the
main points of \cite{hqet:pap1}. 
We regularize the theory on a space-time lattice. In the $1/\mbeauty$ 
part of the Lagrangian,
\bes \label{e:lag1}
    \lnu{1}(x) &=& \sum_i \lcoeff{1}{i}\opi{i}{1}(x)\,, \qquad \\&&
                \opi{1}{1} = \heavyb(-\vecsigma\cdot \vecB)\heavy \,,\quad
                \opi{2}{1} = \heavyb(- {\frac12 \vecD^2 }) \heavy \,
\ees
the chromo-magnetic field $\vecB$ and the 3-d Laplacian 
$\vecD^2$ are then discretized in the standard way. Details
will be irrelevant for our discussion. The coefficients
$\lcoeff{1}{i}$ are functions of the bare coupling $g_0$ 
as well as the mass of the heavy quark. They have to be 
determined such as to match the effective theory to QCD. Matching
at the classical level fixes
\bes
   \lcoeff{1}{1} = 1/\mbeauty + \rmO(g_0^2) = \lcoeff{1}{2} \,.
\ees
Furthermore, we note that also in \eq{e:astat}  
a dimension 4 composite field with coefficient
$\propto 1/\mbeauty + \rmO(g_0^2)$ has to be added on the r.h.s. 
when $1/\mbeauty$ corrections are considered. As an additional
essential ingredient in the formulation of the effective theory
we always expand the formal weight in the path integral,
$\exp\left(\sum_{x} -(\lag{\rm stat}(x)+\lnu{1}(x)+\ldots)\right)$, in 
a power series in $1/\mbeauty$. The correlation functions 
are then defined by
\bes \label{e:pathi}
  \langle \op{} \rangle = \frac1Z \int D[\varphi]\, \op{}[\varphi]\,
                            \exp\left({-a^4\sum_{\vecx}\lag{\rm stat}(\vecx)}\right) \,
		\times \left\{ 1 -  a^4 \sum_x \lnu{1}(x) + \ldots\right\}\, ,
\ees
where $\varphi$ denotes collectively the fields of the theory
and the denominator $Z$ insures $\langle 1 \rangle = 1$. 
The higher order terms 
in the Lagrangian then appear only as insertions into
the correlation functions of the static effective theory. 
The latter is renormalizable by power counting and as a result
also the effective theory truncated 
{\em at any finite order in $1/\mbeauty$ is renormalizable}. 
With higher dimensional operators in the 
exponential, as in NRQCD, this would not be the case. 
For the lattice theory renormalizability is important because
it means that the continuum limit exists and is
independent of the details of the lattice formulation (universality).

\subsection{ 
	Power divergencies} \label{s:powdiv}

The coefficients $\dmstat,\lcoeff{1}{i}$ in \eq{e:statlag} and \eq{e:lag1}
have a regular expansion in the bare coupling $g_0^2$.  
Still, perturbative precision is in general insufficient, 
since operators of higher dimensions
mix with those of lower dimension, e.g.
\bes
  \op{\rm R}^{\rm  d=5} = \sum_{k} z_{k}
                                                  \op{ \it k}^{\rm d=5}
                     + \sum_{k} { c_{k}} \op{ \it k}^{\rm d=4} \,, \quad
                   { c_{k}}  = {c_{k}^{(0)} + c_{k}^{(1)}g_0^2 + \ldots
                                \over { a}     } \,.
\ees
Since the lattice spacing decreases as
$a\sim \exp(-1/(2b_0 g_0^2))$ for small bare gauge coupling $g_0$,  a  truncation 
of the perturbative series leaves terms undetermined which diverge
as the lattice spacing goes to zero. The origin of this
problem is the same as the need for non-perturbative conversion
functions, but the consequence is more drastic due to
the presence of the hard cutoff in the lattice theory. Without 
non-perturbative precision for $\dmstat,\lcoeff{1}{i}$, the continuum limit
does not exist.

\subsection{Matching strategy} \label{s:strategy}

The definition of the effective theory is essentially given by
\eq{e:pathi}, supplemented by a definition of the effective 
composite fields. The only missing piece is a practical
strategy for ascertaining how the parameters in the Lagrangian and in the
effective fields can be determined beyond perturbation theory.
At a given order $n$ in the $1/\mbeauty$-expansion, we denote
the parameters in the effective theory by $c_k$, $k=1,\ldots,\Nn$. Observables,
e.g. renormalized correlation functions or energies are denoted
by $\Phihqet_k(L,M)$ ($\Phiqcd_k(L,M)$) in the 
effective theory (in QCD), with the argument $M$ referring to the 
mass of the heavy quark and $L$ the linear extent of the finite volume. 
It is then sufficient to impose
\bes
  \label{e:match}
   \Phihqet_k(L_0,M) = \Phiqcd_k(L_0,M)\,, \quad k=1,\ldots,\Nn\,.
\ees
to determine all parameters
$\{c_k\,,k=1,\ldots,\Nn\}$ in the effective theory.
Observables used originally to fix 
$\{c_k\,,k=1,\ldots,\Nf\}$,
the parameters of QCD, 
may be amongst these $\Phiqcd_k$.
The matching conditions, \eq{e:match}, define the set
$\{c_k\}$ for any value of the lattice spacing (or equivalently
bare coupling). Here, a typical choice is $L_0 \approx 0.2 \ldots 0.4\,\fm$,
since in such a volume the r.h.s. of the equation
can be evaluated, see \sect{s:tests}.
In practice, the parameters of the effective theory are then determined
at rather small lattice spacings in a range of $a=0.01\,\fm$
to $a=0.04\,\fm$. Large volumes as they are needed to compute
the physical mass spectrum or matrix elements then require
very large lattices ($L/a > 50$). A further step is needed to bridge 
the gap to practicable lattice spacings. A well-defined procedure is 
as follows:
First we assume that all observables $\Phihqet_k(L,M)$ have been made
dimensionless by multiplication with appropriate
powers of $L$. Next, we define step scaling
functions~\cite{alpha:sigma}, $F_k$, by
\bes
\label{e:ssfcont}
   \Phihqet_k(sL,M) = F_k(\{\Phihqet_j(L,M)\,,\,j=1,\ldots,\Nn\})\,,\quad
   k=1,\ldots,\Nn \,,
\ees
where typically one uses scale changes of $s=2$.
These dimensionless functions describe the change of
the complete set of observables $\{\Phihqet_k\}$ under a scaling of
$L\to sL$.
In order to compute them one selects a lattice with a certain resolution
$a/L$. The specification of $\Phihqet_j(L,M)$, $j=1,\ldots,\Nn$, then
fixes all (bare) parameters of the theory. The l.h.s.~of \eq{e:ssfcont}
is now computed,  keeping the bare parameters fixed
while changing $L/a \to L'/a = sL/a$. 
The values for the continuum $F_k$ can then be  be reached by 
extrapolating the resulting lattice numbers to $a/L\rightarrow 0$.

After repeating this step two or three times with $s=2$,
lattice spacings appropriate for infinite volume computations
will be reached.

\subsection{Example: the mass of the b-quark at lowest order} \label{s:mb}

\begin{wrapfigure}{r}{8.0cm}
 \vspace{-0.5cm}
 \hspace{0.5cm}\includegraphics[width=7.5cm]{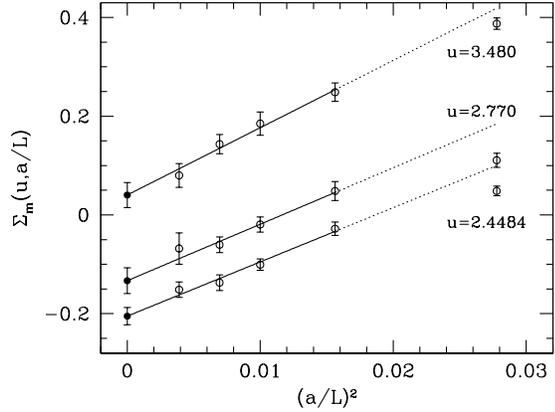}
 \vspace{-0.5cm}
 \hspace{0.5cm} \hfill \caption{
\footnotesize
Extrapolation $\Sigmam(u,a/L) = \sigmam(u) + c\, {a^2 / L^2}$ 
}
\label{f:contextr}
\end{wrapfigure}

For illustration purposes
we consider  a simple 
example here, the computation of the b-quark mass, starting
from the observed B-meson mass. Already at the lowest order in $1/\mbeauty$
the mixing of operators of different dimensions is relevant
in this case: $\heavyb D_0\heavy$ mixes with  $\heavyb \heavy$.
Hence $\dmstat = (c_1 g_0^2 + c_2 g_0^4 + \ldots)/a$,
the coefficient of $\heavyb \heavy$ in the 
Lagrangian \eq{e:statlag}, has to be 
determined (or eliminated) non-perturbatively.  In \eq{e:match}
we have 
$n=0$, $N_0=\nf+1$ and we omit 
the discussion of the choices for $\Phi_k$, $k=1,\ldots,\nf$ which fix
the bare light quark masses and coupling
(both in QCD { and} HQET). 
Obviously any energy in the b-sector will do  
to fix  $\dmstat$.
We choose\cite{hqet:pap1,hqet:pap2} \footnote{
In practice $ \meff$ is replaced by the spin averaged energy \cite{hqet:pap1,hqet:pap2}.}
\bes \label{e:defgam}
  \meff(L,M) = \frac{1}{2a} \ln\Big[\,\fa(x_0-a)/\fa(x_0+a)\,\Big] \qquad
  \left(\mbox{$x_0/L$ fixed}\right)\,
\ees
and we require \eq{e:match} where for $k=\nf+1$ we identify
\bes
  \Phiqcd_{\nf+1}(L,M)\equiv L\,\meff(L,M)\,,\quad   
 \Phihqet_{\nf+1}(L,M)\equiv L\,(\meffstat(L) + m) \,,
\ees
Here $m$ represents the quark mass that was removed from all
energies in defining the effective theory such that the $m\to\infty$
limit exists and $\meffstat$ refers to \eq{e:defgam} at the lowest order 
in $1/M$. The relevant part of \eq{e:ssfcont} can then 
be written in the simple form,
\bes
\label{e:ssfmass}
    \Phihqet_{\nf+1}(2L,M) &=&   2 \Phihqet_{\nf+1}(L,M) + \sigmam\left(\gbar^2(L)\right)\,,
    \\
    \sigmam\left(\gbar^2(L)\right) &\equiv& 2L\, [\,\meffstat(2L) - \meffstat(L)\,]\,.
\ees
In $\sigmam$ the divergent $\dmstat$ (as well as the mass shift $m$) cancel.
It is independent of the mass. 
We now see immediately that 
\def\text#1{\mbox{#1}}
\bes \label{e:master}
  \mB 
      = \underbrace{\Estat - \meffstat(L_2)}_{a\to0 \text{ in HQET}} +
        \underbrace{\meffstat(L_2)-\meffstat(L_0)}_{a\to0 \text{ in HQET}}
       +\underbrace{\meff(L_0,{ \Mb})}_
		{a\to0 \text{ in QCD} } 
       +\rmO(\Lambda^2/\Mb) \,.
\ees 
Here, $\Estat = \lim_{L\to\infty} \meffstat(L)$ is the infinite volume
energy of a B-meson in static approximation. 
It is often called the static binding energy. 
The whole strategy is illustrated in \fig{f:mbstrat}. 
As indicated in \eq{e:master}, the continuum
limit can be taken in each individual step; a numerical example
is shown in \fig{f:contextr}.

\newcommand{\ftext}[1]{\fbox{ {#1} }}
%

\newcommand{\cred}{}
\newcommand{\cblu}{}
\newcommand{\cmag}{}
\newcommand{\cgre}{}
\newcommand{\cbla}{}

\newcommand{\mgt}{\cmag}

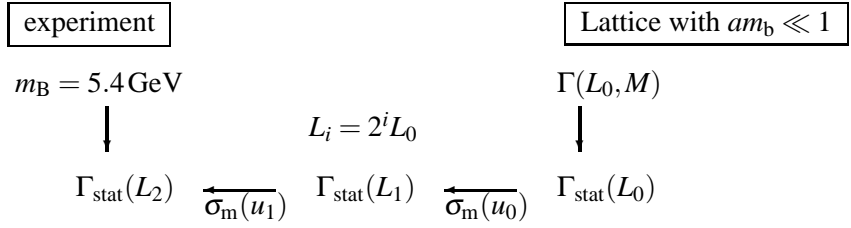
\begin{figure}[htb]
\vspace{5.5cm}

\begin{picture}(8,80)(0,0)
\small
\hspace{-5.5cm}
  \unitlength 0.4cm
  \put(2,6){\ftext{experiment}}            \put(20.5,6){\ftext{Lattice with 
$a\mbeauty\ll 1$}} 
  \put(2,4){ $\mB=5.4\,\GeV$}    \put(20,4){ $\meff(L_0,M)$} 
  \linethickness{0.3mm}\cgre\put(5.3,3.5){\vector(0,-1){1.5}}
  \linethickness{0.3mm}\cgre\put(21,3.5){\vector(0,-1){1.5}}
  \linethickness{0.3mm}\cbla
  \put(4,0.5){ $\meffstat(L_2)$}
  \put(12,0.5){ $\meffstat(L_1)$}
  \put(20,0.5){ $\meffstat(L_0)$}
  \put(18.9,0.7){\vector(-1,0){2.4}}
  \put(10.9,0.7){\vector(-1,0){2.4}}
  \put(16.5,-0.1){$\sigmam(u_0)$}
  \put( 8.5,-0.1){$\sigmam(u_1)$}
  \put(12,2.5){\small $L_i = 2^i L_0$}
\end{picture}
 \caption{
 Connecting experimental observables to renormalized HQET.
 $\meffstat$ is a renormalized quantity in HQET and $\sigmam(\gbar^2(L))$
 connects $\meffstat(L)$ and $\meffstat(2L)$. 
 }
 \vspace{-0.5cm}
 \label{f:mbstrat} 
\end{figure}

After obtaining all pieces in \eq{e:master},
the equation is numerically solved for $z_\beauty=\Mbeauty L_0$.
Since also the size of  $L_0$ in units of $r_0\approx0.5\,\fm$ \cite{pot:r0} is known, 
one can 
quote (remember $ \mbeauty$ is in the $\msbar$ scheme at scale $ \mbeauty$)
\bes
 r_0 \Mbeauty = 16.12(24)(15)\; \to\;\Mbeauty = 6.36(10)(6)\,\GeV \,,\quad 
 \mbeauty = 4.12(7)(4)\,\GeV \,.
\ees
We emphasize that this result is in the quenched approximation but includes
the lowest non-trivial order in $1/\mbeauty$. An estimate of the associated
$\rmO(\Lambda^2/ \Mbeauty)$ uncertainty is {\em not} included in the 
errors shown. Our discussion mainly serves to illustrate the potential of
the approach in an example where the
power divergent mixing needed to be solved non-perturbatively.

\section{Perspectives} \label{s:persp}

%
Non-perturbative HQET at the leading order in $1/m$ has reached a satisfactory status,
with the b-quark mass \cite{hqet:pap1} and the B$_\strange$-meson decay constant
\cite{zastat:pap3,stat:letter}
known in the continuum limit of the quenched approximation. Their 
precision can and will still be improved. Applying these methods 
to the theory with dynamical fermions is straight forward; ''only''
the usual problems with the light quarks have to be solved.
By themselves such lowest order results are not expected to
have an interesting precision for phenomenological applications,
but certainly they can constrain the large mass behavior computed
with other methods \cite{fb_wupp,reviews:beauty,reviews:hartmut97,El-Khadra:1998hq,
Aoki:1998ji,Bernard:1998xi,heavylight:Bec98,AliKhan:2000eg,Bowler:2000xw,
Lellouch:2000tw,lat01:ryan,mb:roma2,fb:roma2c}.

More interestingly, also $1/\mbeauty$ corrections can, in principle, be
computed in the effective theory. Here, details of the 
necessary numerical steps have not yet been implemented but the very
first tests have been encouraging \cite{lat04:stephan}.
Another relevant technical advance has been the realization
that a change of the regularization details allows to
achieve much better statistical errors in HQET, while 
keeping the discretization errors small \cite{stat:letter}.
In summary, we believe that all ingredients exist which are
needed to apply HQET beyond the leading
order in $1/\mbeauty$.


\begin{theacknowledgments}
  I would like to thank my colleagues S. D\"urr, M. Della Morte, 
  A. J\"uttner, J. Rolf, and in particular J. Heitger for a 
  pleasant collaboration in developing non-perturbative HQET. 
  I am grateful to H. Simma for useful comments on this manuscript.
\end{theacknowledgments}



\bibliographystyle{aipproc}   

\bibliography{cag}

\IfFileExists{\jobname.bbl}{}
 {\typeout{}
  \typeout{******************************************}
  \typeout{** Please run "bibtex \jobname" to optain}
  \typeout{** the bibliography and then re-run LaTeX}
  \typeout{** twice to fix the references!}
  \typeout{******************************************}
  \typeout{}
 }

\end{document}